\documentclass{article}[11pt]
\usepackage{amssymb,amsmath,amsthm}
\usepackage{multicol,graphicx}
\newtheorem{theorem}{Theorem}

\newtheorem{fact}{Fact}
\newtheorem{lemma}{Lemma}

\newtheorem{definition}{Definition}

\newcommand{\globalg}{\tt{Density}}
\newcommand{\alg}[2]{\tt{Local Density}({#1},{#2})}

\newcommand{\norm}[2]{\|{#2}\|_{#1}}
\newcommand{\ip}[2]{\left \langle {#1},{#2} \right \rangle}
\newcommand{\set}[2]{X^{#1}_{#2}}
\newcommand{\settwo}[2]{Y^{#1}_{#2}}

\newcommand{\edges}{\operatorname{e}}
\newcommand{\density}{d}
\newcommand{\truncate}[1]{\operatorname{prune_{\epsilon_{#1}}}}
\newcommand{\round}{\operatorname{round}}
\newcommand{\Arest}[1]{A_{(#1)}}
\newcommand{\good}[2]{\operatorname{#2_{#1}}}

\newcommand{\supp}{{\rm Support}}

\newcommand{\ignore}[1]{}
\begin{document}
\title{A Local Algorithm for Finding Dense Subgraphs}
\author{
Reid Andersen} 
 \maketitle

 \begin{abstract} We present a local algorithm for finding dense
     subgraphs of bipartite graphs, according to the 
     definition of density proposed by Kannan and Vinay.  Our algorithm takes as
     input a bipartite graph with a specified starting vertex, and
     attempts to find a dense subgraph near that vertex.  We prove
     that for any subgraph $S$ with $k$ vertices and density $\theta$,
     there are a significant number of starting vertices within $S$
     for which our algorithm produces a subgraph $S'$ with 
     density $\Omega(\theta / \log n)$ on at most
     $O(\Delta k^{2})$ vertices,
     where $\Delta$ is the maximum degree.  The running time of
     the algorithm is $O(\Delta k^{2})$, independent of the number of
     vertices in the graph.  \end{abstract}
 
 \section{Introduction} 
Identifying dense subgraphs has become an
 important task in the analysis of large networks, and a collection of
 dense subgraphs may reveal a wealth of information about a
 graph.  In particular, dense subgraphs often form the
 cores of larger communities or clusters in the graph~\cite{trawling}. 

Kannan and Vinay~\cite{Kannan} introduced a notion of
density that is well-suited to bipartite graphs representing 
incidence matrices. 
As an example, consider a bipartite graph describing the 
incidences between a set of groups
$\mathcal{G}$ and a set of group members $\mathcal{M}$.
The density of the subgraph induced by 
a set of groups $S \subseteq \mathcal{G}$ and a set of members $T \subseteq \mathcal{M}$,
is defined to be 
\[d(S,T) =
\frac{e(S,T)}{\sqrt{|S|} \sqrt{|T|}},\]
which is the total number of incidences between 
the groups and members in the subgraph,
divided by the geometric mean of the number of groups and
number of members in the subgraph.
There are a variety of efficient algorithms for
finding a subgraph with nearly optimal density according to this 
definition.
Kannan and Vinay gave a spectral algorithm that 
produces from the largest
eigenvector of $A$ 
a subgraph whose density is within an $O(\log n)$
factor of optimal.  Charikar~\cite{Charikar} showed that a subgraph
with optimal density can be identified in polynomial time by solving a
linear program, and also gave a greedy algorithm
that produces a 2-approximation of the densest subgraph in linear
time.

In this paper, we present a local algorithm for finding
dense subgraphs.
Our algorithm takes as input a graph with a
specified starting vertex, and attempts to find a dense subgraph near
that vertex.  We prove the following local approximation guarantee for
our algorithm: for any subgraph $H$ with density $\theta$,
there are a significant number of starting vertices
within $H$ for which our algorithm produces a subgraph with density
$\Omega(\theta / \log n)$.  The running time of the algorithm is
$O(\Delta k^{2})$, where $k$ is the number of vertices in $H$, and
where $\Delta$ is the maximum degree in the graph. 

There are two principal tasks that our local algorithm can perform
which, to our knowledge, can not be accomplished by other known algorithms
for the densest subgraph problem.
The first is to find a dense subgraph near a vertex of interest, while
examining only a portion of the entire graph.  The second is to find
many small dense subgraphs in parallel, which we can accomplish by applying 
the local algorithm at many different
starting vertices.  In addition, our algorithm provides an upper bound on the size
of the subgraph it produces, which might make it a useful theoretical
tool for producing a dense subgraph of a specified size.

To analyze our algorithm, we build upon the spectral techniques
developed by Kannan and Vinay, exploiting the close
relationship between the densest subgraph of a graph and the largest
eigenvalue of the graph's adjacency matrix.  We define a deterministic
process called the  `pruned growth process', which produces a
sequence of vectors, and show that by computing those vectors we can
identify a subgraph with high density.  We show that these vectors can be 
rounded at each step to ensure that the number of nonzero elements 
is small, which
decreases the time required to compute them.  A similar type of local
approximation algorithm has been developed for the related problem of
graph partitioning \cite{Spielman, Andersen}.  
The densest subgraph
problem is the second problem for which this type of local spectral
algorithm has been developed.

In Section~\ref{S:density}, we state the definition of density
introduced by Kannan and Vinay, compare this definition with others
that have appeared in the literature, and survey known algorithms for
the densest subgraph problem.  In Section~\ref{S:process}, we define
the `pruned growth process'.  In Section~\ref{S:main}, we state our
local algorithm and analyze its running time and approximation
guarantee.  In Section~\ref{S:global}, we describe an efficient global approximation 
algorithm for the densest subgraph problem, which will follow
easily from our work in the previous sections.

\section{Preliminaries and Related Work}\label{S:density} Let
$G=(V,E)$ be an undirected bipartite graph with adjacency matrix $A$,
and let $L$ and $R$ be the left and right sides of a fixed bipartition.
The edges of the graph may be weighted, in which
case the entry $A_{i,j}$ is the weight of edge $\{i,j\}$.  For any two sets $S \subseteq L$ and $T
\subseteq R$,
we let $(S,T)$ denote the induced bipartite subgraph of $G$ on the set
of vertices $S \cup T$,
and we define $\edges(S,T)$ to be the sum of the weights of
the edges between $S$ and $T$.  We will sometimes use the inner
product notation $\edges(S,T) = \ip{1_{S}A}{1_{T}}$, where $1_{S}$ is
the indicator function for membership in $S$.  We define the {\em
support} of a vector $x$ to be the set of vertices on which $x$ is
nonzero. 

We will identify induced subgraphs of $G$ which are dense according the following
definition, which was introduced by Kannan and Vinay~\cite{Kannan}.
\begin{definition}\label{D:density} For any induced subgraph $(S,T)$,
    we define $$\density(S,T) =
    \frac{\edges(S,T)}{\sqrt{|S|}\sqrt{|T|}}.$$ We define
    $\density(A)$ to be the maximum value of $\density(S,T)$ over all
    induced subgraphs.  \end{definition} Our algorithm may also be applied to
    an arbitrary directed graph, using the following trick.  Given a
    directed graph with vertex set $X$, define a bipartite graph where
    $L=R=X$. For each edge $x \rightarrow y$ in the directed graph,
    place an undirected edge between the copy of $x$ in $L$ and the
    copy of $y$ in $R$.  

\subsection{Related work}

A different definition of density was considered in \cite{Goldberg,GGT,Charikar}.  
\begin{definition} 
    Let $G=(V,E)$ be an undirected graph (not necessarily bipartite).
For any set $S\subseteq V$, we define
$$g(S) 
= \frac{\edges(S,S)}{|S|}.$$
We define $g(A)$ to be the maximum value of $g(S)$ over all subsets of $V$.
\end{definition}

Both $d(A)$ and $g(A)$ can be computed exactly in polynomial time.
Goldberg showed that a set $S$ achieving $g(S)=g(A)$ can be found
using maximum flow computations \cite{Goldberg}.  Such a set can also
be
found using the parametric flow algorithm of Gallo, Grigoriadis, and
Tarjan \cite{GGT}.  Charikar showed that a subgraph $(S,T)$ achieving
$\density(S,T)=\density(A)$ can be found by solving a linear program
\cite{Charikar}.

Charikar gave greedy 2-approximation algorithms for both $g(A)$ and
$\density(A)$ \cite{Charikar}.  The running time of these algorithms
is $O(m)$ in an unweighted graph, and $O(m + n\log n)$ in a weighted
graph.  Kannan and Vinay gave a spectral approximation algorithm for
$\density(A)$, which produces a subgraph $(S,T)$ with density $d(S,T) =
\Omega(\density(A)/\log n)$ from the top singular vectors of $A$.  

The closely related densest $k$-subgraph problem
is to identify the subgraph with the largest number of edges among all
subgraphs of exactly $k$ vertices.  This problem is considerably more
difficult,
and there is a large gap between the
best approximation algorithms and hardness results known for
the problem (see \cite{feigeseltser,feige}).

\subsection{Comparison of $d(S,T)$ and $g(S)$}
It is easier to compare the two objective functions
$d(S,T)$ and $g(S)$ if we restrict $g(S)$ to bipartite graphs.  In this case, $g(S)$ takes the following form.
\begin{definition}
For any subgraph $(S,T)$, we define
$$g(S,T) 
= \frac{\edges(S,T)}{|S|+|T|}.$$
We define $g(A)$ to be the maximum value of $g(S,T)$ over all induced subgraphs.
\end{definition}
The two objective functions $\density(S,T)$ and $g(S,T)$ are far apart when $S$ and $T$ have very different sizes.
The quantities $\density(A)$ and $g(A)$ can also be far apart.  In the complete bipartite graph $K_{a,b}$, we have 
\mbox{$\density(A) = \sqrt{ab}$}, 
    while \mbox{$g(A) = ab/(a+b)$}.
In the case where $a=1$,
we have \mbox{$\density(A) = \sqrt{b}$} while 
\mbox{$g(A) = b/(b+1) \sim 1$}.

The relative merits of
$d(S,T)$ and $g(S)$ as objective functions for density were discussed in \cite{Charikar,Kannan}.
In this paper, we consider $\density(S,T)$
because it is more amenable to approximation by spectral algorithms
than $g(S)$,
not because we prefer it as an objective function.
The largest eigenvalue of the adjacency matrix $A$ 
is closely related to $\density(A)$.
We know of no similar result for $g(A)$,
and we do not know how to produce a local algorithm 
for the objective function $g(S)$.

\section{The pruned growth process}\label{S:process}
We now define the deterministic process that will be the basis for our
local algorithm.
The process generates a sequence of vectors $x_{0}, \dots, x_{T}$ from a starting vector $x_{0}$.  The main operation performed at each step 
is multiplication by the adjacency matrix $A$, as in the power method.  
The resulting vector is then rounded by making each entry a power of 2, and then pruned by setting to zero each entry whose value is below a certain threshold.
These steps reduce the number of possible values in the vector and reduce the size of the support, minimizing the amount of computation required.  

\begin{definition}\label{ops}
Given a vector $z$, we define
$\round(z)$ to be the vector obtained 
by rounding each entry of the vector 
$z$ up to the nearest power of 2, 
\[ [\round(z)](u) =
    2^{i} \mbox{, where $i$ is the smallest integer such that 
    $2^{i} \geq z(u)$.}
\]

Given a vector $z$ and a nonnegative real number $\epsilon$,
we define
$\truncate{}(z)$ to be the vector obtained by setting to zero
any entry of $z$ whose value is at most $\epsilon \norm{}{z}$,
\[ [\truncate{}(z)](u) =
\begin{cases}
    z(u) \qquad \mbox{ if } 
    z(u) > \epsilon \norm{}{z},\\ 
    0 \qquad \mbox{ otherwise.}  
\end{cases}
\]
\end{definition}
\begin{definition}\label{process}
    Given a starting vector $x_{0}$ with entries from $\{0,1\}$,
    and a sequence of real numbers $\epsilon_{t} \in [0,1]$,
    we define the {\em pruned growth process} to be the sequence of vectors $x_{0}, \dots, x_{T}$
    defined by the following rule:
    \begin{align*}
        x_{t+1} &= \truncate{t+1}(\round(x_{t}A)).
    \end{align*}
Notice that each entry of $x_{t+1}$ is either zero or a power of two.
\end{definition}

\begin{definition} Given the vectors $x_{0}, \dots, x_{T}$ of the pruned growth
process, 
we define $\set{t}{i}$ to be the set of vertices where $x_{t}(v)=2^{i}$,
 and define $\settwo{t}{i}$ to be the set of vertices where $\round(x_{t}A)(v)=2^{i}$.  \end{definition}

 We will eventually show that a subgraph with high density can be found
 whenever the norms $\norm{}{x_{i}}$ of the pruned growth process vectors
 grow quickly.
The following lemma shows that if none of the subgraphs 
$(\set{t}{i},\settwo{t}{j})$ has high density, then $\| x_{t}\|$ is not much larger than $\| x_{t-1}\|$.
\begin{lemma}\label{growthbound}
If $\density(\set{t}{i},\settwo{t}{j}) \leq \theta$ for all $i,j$,
then,
$$
\| x_{t+1} \| 
\leq \| \round(x_{t}A) \| 
\leq 2 \theta \|x_{t} \| 
    \log\frac{2\Delta}{\epsilon_{t}}.
$$
 \end{lemma}
 \begin{proof}
 We will write $\| \round(x_{t}A) \|^{2}$
 in terms of the 
 densities $\density(\set{t}{i},\settwo{t}{j})$.
 \begin{align*}
     \| \round(x_{t}A) \|^{2}
     &= \ip{\round(x_{t}A)}{\round(x_{t}A) } \\
     &\leq \ip{2x_{t}A}{\round(x_{t}A)} \\
     &= \ip{2\sum_{i} 2^{i} 1_{\set{t}{i}}A}
     {\sum_{j} 2^{j} 1_{\settwo{t}{j}}} \\
     &= 2\sum_{i,j} 2^{i}2^{j} \ip{1_{\set{t}{i}}A}{1_{\settwo{t}{j}}}  \\
     &= 2\sum_{i,j} 2^{i} 2^{j} \density(\set{t}{i},\settwo{t}{j}) 
     \sqrt{|\set{t}{i}|}\sqrt{|\settwo{t}{j}|} \\
     &\leq 2 \theta 
     \left( \sum_{i} \sqrt{|\set{t}{i}|} 2^{i} \right) 
     \left( \sum_{j} \sqrt{|\settwo{t}{j}|}2^{j} \right).
 \end{align*}
 In the sum above, we need only sum over those $i$ 
 where $\set{t}{i}$ is nonempty.  There are at most 
 $\log{\frac{1}{\epsilon_{t}}}$ such values,
 because every nonzero value in $x_{t}$ is
 at most $\norm{}{x_{t}}$ and at least $\epsilon_{t} \norm{}{x_{t}}$.  
 We now apply the Cauchy-Schwarz inequality to show 
 \[
 \sum_{i} \sqrt{|\set{t}{i}|} 2^{i} 
 \leq 
 \left(\sum_{i} |\set{t}{i}| 2^{2i} \right)^{1/2} 
 \left(\sum_{i} 1 \right)^{1/2} 
 \leq
 \|x_{t}\|
 \sqrt{\log\frac{1}{\epsilon_{t}}} 
 .\]
 Similarly, we need only sum over those $j$ 
 where $\settwo{t}{j}$ is nonempty.  There are at most 
 $\log{\frac{2\Delta}{\epsilon_{t}}}$ such values,
 because every nonzero value of 
 $\round(x_{t}A)$ is at most
 $\norm{ }{\round(x_{t}A)}
 \leq 2\Delta \norm{}{x_{t}}$,
 and at least $\epsilon_{t} \norm{}{x_{t-1}}$.  
 We apply the Cauchy-Schwarz inequality again to show
 \[
 \sum_{j} \sqrt{|\settwo{t}{j}|} 2^{j} 
 \leq 
 \left(\sum_{j} |\settwo{t}{j}| 2^{2j} \right)^{1/2} 
 \left(\sum_{j} 1 \right)^{1/2} 
 \leq
 \|\round(x_{t}A)\|
 \sqrt{\log\frac{2\Delta}{\epsilon_{t}}}.
 \]
 Then, \begin{align*}
     \| \round(x_{t}A) \|^{2} 
     &\leq 2 \theta \|x_{t} \| \|\round(x_{t}A)\| 
     \sqrt{\log\frac{1}{\epsilon_{t}}}
     \sqrt{\log\frac{2\Delta}{\epsilon_{t}}}\\
     &\leq 2 \theta \|x_{t} \| \|\round(x_{t}A)\| 
     \log\frac{2\Delta}{\epsilon_{t}}.
 \end{align*}
 The lemma follows. 
 \end{proof}

 \section{Local approximation algorithm}\label{S:main} In this
 section, we will state and analyze a local algorithm for finding
 dense subgraphs.  The input to the algorithm is a graph, along with a
 {\em starting vertex} $v$ and a {\em target size} $K$.  We will prove
 that the running time of the algorithm depends mainly on the target
 size $K$, and is independent of the number of vertices in the graph.
 We will prove that for any subgraph $(S,T)$, there are a
 significant number of starting vertices in $S$ for which the
 algorithm produces a subgraph whose density is 
 within an $O(\log n)$ factor of $d(S,T)$.

 \noindent
 \framebox{
 \begin{minipage}{.95\textwidth}
     {\noindent $\alg{v}{K}$}\\
     Input:  A vertex $v$ and a target size $K$.\\
     Output:  A subgraph $(X,Y)$.
     \begin{enumerate}
         \item Let $x_{0} = 1_{v}$, let $T = \log(\sqrt{2|K|})$, and 
             let $\epsilon_{t} = \frac{1}{8K}2^{-t}$.

         \item
             Compute the vectors $x_{0}, \dots, x_{T}$ of the pruned growth process. 
         \item 
             Compute \mbox{$\density(\set{t}{i},\settwo{t}{j})$} for each pair $i$, $j$ and each time $t<T$. 
         \item
             Output the subgraph 
             $(\set{t}{i},\settwo{t}{j})$
             with the highest density.
     \end{enumerate}
 \end{minipage}
 }

 \begin{theorem}\label{localalg}
     Let $(S,T)$ be a subgraph such that \mbox{$\density(S,T) \geq 2\theta$}.  Then there exists a set $\good{\theta}{S}\subseteq S$,
     with the following properties.
     \begin{enumerate}
         \item $e(\good{\theta}{S},T) \geq e(S,T)$,
         \item If $v \in \good{\theta}{S}$
             and \mbox{$K \geq \max(|S|,|T|)$}, then 
             $\alg{v}{K}$ outputs a subgraph $(X,Y)$ such that
             \[\density(X,Y) \geq \frac{\theta}{8 \log 16\Delta K}
             = \Omega(\frac{\theta}{\log n}).\] 
     \end{enumerate}
 \end{theorem}
 
 \begin{theorem}\label{runningtime}
     $\alg{v}{K}$ runs in time $O(\Delta K^{2})$.
 \end{theorem}
 The proofs of Theorems~\ref{localalg} and \ref{runningtime} 
 are given in section~\ref{SS:proofs}.

 \subsection{Lower bounds on growth within a dense subgraph}\label{SS:lower}
 The main step in analyzing the algorithm {\tt LocalDensity} is to
 prove a lower bound on the growth of the norms $\norm{ }{x_{t}}$.
 We will use the fact that the maximum density $\density(A)$ gives a lower bound on the largest eigenvalue of $A$.
 \begin{fact}\label{eigenvaluefact}
     Let $A$ be the adjacency matrix of an undirected graph,
     and let $\lambda$ be the largest eigenvalue of $A$. 
     Then, $\lambda \geq d(A)$.   Furthermore, there is an eigenvector
     $\phi$ with eigenvalue $\lambda$ whose entries
     are nonnegative.
 \end{fact}
 \begin{proof}
     To prove that $\lambda \geq d(A)$, notice that for any sets $S \subseteq L$ and $T \subseteq R$, 
     \[\lambda \geq \max_{x,y}\frac{\ip{xA}{y}}{\norm{ }{x}\norm{ }{y}} 
     \geq 
     \ip{\frac{1_{S}}{\sqrt{|S|}} A}{\frac{1_{T}}{\sqrt{|T|}}} 
     =\frac{\edges(S,T)}{\sqrt{|S|}\sqrt{|T|}}
     = \density(S,T) .\]
     It is not hard to see that if $\phi$ 
     is an eigenvector with eigenvalue $\lambda$,
     then the vector whose entries are the absolute values of the
     entries of $\phi$ is also an eigenvector with
 eigenvalue $\lambda$.
 \end{proof}

 The fact above implies the lower bound $\norm{ }{x_{0}A^{t}} \geq
 \ip{\phi}{x_{0}}d(A)^{t}$, which depends on the maximum density
 $d(A)$.  To analyze the local algorithm, we will give
 a lower bound that depends on the density of a particular subgraph
 $(S,T)$ containing the starting vertex.  Specifically, we will show
 that for many vertices in the set $S$, we can give a bound of the
 form $\norm{ }{x_{0}A^{t}} = \Omega(d(S,T)^{t})$ with a not-too-small
 constant term.  We will do so by considering how the pruned growth
 process would behave if it were restricted to the induced subgraph 
 $(S,T)$. 

 \begin{definition}
     For any induced subgraph $(S,T)$,
    we define $\Arest{S,T}$ to be the restriction of the 
    adjacency matrix $A$ to $(S,T)$, 
     \[
     \Arest{S,T}(x,y) = \begin{cases}
         \mbox{ $A(x,y)$ if  $x \in S$ and $y \in T$,
         or if $x \in T$ and $y \in S.$}\\ 
         \mbox{ $0$ otherwise.}\\ 
     \end{cases}
     \]
 \end{definition}

 The following lemma identifies, for any subgraph $(S,T)$, a set of starting vertices for which we can give a good lower bound on the 
 norms $\norm{ }{x_{t}}$.  This set of good starting vertices touches at least 
 half of the edges in the induced subgraph $(S,T)$.
 \begin{lemma}\label{goodstartlemma}
     If $(S,T)$ is a subgraph such that $\density(S,T) \geq 2\theta$,
     then there exists a subset
     $\good{\theta}{S} \subseteq S$ with the following properties:
     \begin{enumerate}
         \item $e(\good{\theta}{S},T) \geq e(S,T)/2$
         \item For each $v \in \good{\theta}{S}$, 
             there is a nonnegative unit vector $\psi$ such that
             \begin{enumerate}
                 \item $\supp(\psi) \subseteq S\cup T$,
                 \item $\psi A \geq \theta \psi$,
                 \item $\psi(v) \geq \frac{1}{\sqrt{2|S|}}.$
             \end{enumerate}
     \end{enumerate}
 \end{lemma}

 \begin{proof}
     Let $\good{\theta}{S}$ be the largest subset of $S$ for which property (2)
     holds, and consider the set $S' = S \setminus \good{\theta}{S}$.
     If $\good{\theta}{S}$ does not satisfy property (1),
     then 
     \[ e(S',T) = e(S,T) - e(\good{\theta}{S},T)
     \geq
     \frac{e(S,T)}{2},\]
     and so $\density(S',T) \geq \density(S,T)/2 \geq \theta$.

     Let $\lambda$ be the largest eigenvalue of $\Arest{S',T}$.
     We know from fact~\ref{eigenvaluefact} that
     there is an eigenvector $\psi$ of $\Arest{S',T}$
     whose entries are all nonnegative,
     and whose corresponding eigenvalue $\lambda$ 
     satisfies
     \[\lambda \geq \density(S',T) \geq \theta.\]
     It is easy to see that $\psi$ satisfies properties (a) and (b).
     We will now identify a vertex in $S'$ for which 
     $\psi(v) \geq 1/\sqrt{2|S|}$.  This will imply that $v$ is in $\good{\theta}{S}$, which will show that $\good{\theta}{S}$ must
     satisfy property (1), and thus complete the proof.  

     Let $\psi_{S'}$ and $\psi_{T}$ be the projections of $\psi$ onto $S'$ and $T$, and observe that 
     \mbox{$\|\psi_{S'}\| = \|\psi_{T}\| = \frac{1}{\sqrt{2}}$}.  This is true because \mbox{$\psi_{S'}\Arest{S',T} = \lambda \psi_{T}$},
     which implies that 
     \mbox{$ \lambda \|\psi_{S'}\| \geq \| \psi_{S'}\Arest{S',T} \| = \lambda \| \psi_{T} \|$}.
     There must at least one vertex $v$ in $S'$ which satisfies
     $\psi(v) \geq 1/\sqrt{2|S'|}$,
     since otherwise we would have
     $\|\psi_{S'}\|^{2} < 1/2$.
 \end{proof}

 \subsection{Analysis of the local algorithm}\label{SS:proofs}
 \begin{proof}[{\bf Proof of Theorem~\ref{localalg}}]
     We will prove that for each vertex $v$ in the set
     $\good{\theta}{S}$,
     which was described in Lemma~\ref{goodstartlemma}, the algorithm 
     $\alg{v}{K}$ outputs a subgraph with density at least
     $\theta/8L$, 
     where 
     $L = \log(2\Delta/\epsilon_{0}) \leq (\log 16\Delta K)$,
     provided that
\mbox{$K \geq \max(|S|,|T|)$}.
The theorem will follow.

     Let $x_{0}, \dots, x_{T}$ be the pruned growth process vectors computed by the algorithm.  
     We will assume that the algorithm does not find a subgraph with the desired density, and derive a contradiction.  That is, we assume that for each $i$, $j$, and each time $t < T$, 
    we have \mbox{$\density(\set{t}{i},\settwo{t}{j}) < \frac{\theta}{8L}$}.
     Under this assumption, Lemma~\ref{growthbound} shows that 
     for every $t \leq T$,
     \begin{align*}
         \| x_{t+1} \| 
         &\leq \| \round(x_{t}A) \| \\
         &< 
         \left(2 \log\frac{2\Delta}{\epsilon_{t}} \right)
         \left(\frac{\theta}{8L} \right) \|x_{t} \|\\
         &\leq \left(\frac{\theta}{4}\right)\|x_{t} \|.
     \end{align*}
     Since $\|x_{0}\| =1$, this implies
     \begin{equation}\label{growthboundeq}
         \| x_{t} \| 
         \leq \| \round(x_{t-1}A) \| 
         < \left(\frac{\theta}{4}\right)^{t}\quad
         \mbox{for every $t \leq T$.}
     \end{equation}

     Since $v \in \good{\theta}{S}$,
     there exists a nonnegative vector $\psi$ such that 
     \mbox{$\psi A \geq \theta \psi$}, such that \mbox{$\supp(\psi) \subseteq S \cup T$}, and such that \mbox{$\psi(v) \geq \frac{1}{\sqrt{2|S|}}$},
     as stated in Lemma~\ref{goodstartlemma}.
     We will prove the following lower bound
     on the inner product of $x_{t}$ with $\psi$.
     \begin{equation}\label{lowerboundeq}
         \ip{x_{t}}{\psi} \geq  \frac{1}{\sqrt{2|S|}}(\theta/2)^{t}
         \quad \mbox{for every $t \leq T$.}
     \end{equation}
     When we prove equation~\eqref{lowerboundeq},
     it will contradict equation~\eqref{growthboundeq}
     when $t=T=\log(\sqrt{2|S|})$, and we will be done.

     We will prove that equation~\eqref{lowerboundeq} holds
     by induction.
     We know it holds for $t=0$.  
   The only difficulty in the induction step is
   to bound the effect of the pruning step
     on the projection of $x_{t}$ onto $\psi$.  We define $r_{t}$ to be the vector that is removed during the pruning step.
     \begin{align*}
         r_{t} 
         &= \round(x_{t-1}A) - x_{t}\\
         &= \round(x_{t-1}A) - \truncate{t}(\round(x_{t-1}A)).\\  
     \end{align*}
     The value of $r_{t}$ at any given vertex is at most
     $\epsilon_{t} \norm{}{\round(x_{t-1}A)}.$
     Since the support of $\psi$ is contained in $S \cup T$, and the support of 
     $r_{t}$ is contained in either $L$ or $R$, the intersection of the two supports contains at most $\max(|S|,|T|)$ vertices.  The inner product of $r_{t}$ and $\psi$ can then be bounded as follows.
     \begin{align*}
         \ip{r_{t}}{\psi} 
        &\leq \epsilon_{t} \norm{}{\round(x_{t-1}A)}
        \sqrt{\supp(r_{t}) \cap \supp(\psi)}\\
        &\leq \epsilon_{t} \norm{}{\round(x_{t-1}A)}\sqrt{K}.\\
     \end{align*}
     We can now bound $\ip{x_{t}}{\psi}$ in terms of $\ip{x_{t-1}}{\psi}$.
     \begin{align*}
         \ip{x_{t}}{\psi} 
         &= \ip{\round(x_{t-1}A) - r_{t}}{\psi}\\
         &= \ip{\round(x_{t-1}A)}{\psi} - \ip{r_{t}}{\psi}  \\
         &\geq  \theta\ip{x_{t-1}}{\psi} - \epsilon_{t} 
        \norm{}{\round(x_{t-1}A}) \sqrt{K}.
     \end{align*}
     We now assume that the induction hypothesis holds for $t-1$,
     which means
     $\ip{x_{t-1}}{\psi} \geq  (1/\sqrt{2|S|})(\theta/2)^{t-1}$.
     Recall that we have assumed for the sake of contradiction that
     $\| x_{t} \| \leq \|\round(x_{t-1}A)\| < \left(\theta/4\right)^{t}$. 
     We will now show that the induction hypothesis holds for $t$.
     \begin{align*}
         \ip{x_{t}}{\psi} 
         &\geq  
         \left(
         \frac{\theta}{\sqrt{2|S|}}
         \left(\frac{\theta}{2}\right)^{t-1}
         \right) 
         - 
         \left(\epsilon_{t} \sqrt{K} 
         \left(\frac{\theta}{4}\right)^{t}\right)\\
         &\geq  \left(\frac{\theta}{2}\right)^{t} 
         \left(
         \frac{2}{\sqrt{2|S|}} - 4 \epsilon_{t} 2^{-t}\sqrt{K} \right)\\
         &\geq  \left(\frac{\theta}{2}\right)^{t} \frac{1}{\sqrt{2|S|}}.
     \end{align*}
     The last step follows because we have set $\epsilon_{t}$ so that 
     \[
     \epsilon_{t} 
     = \frac{2^{-t}}{8K}.
     \]
     This completes the proof.
 \end{proof}
 
 \begin{proof}[{\bf Proof of Theorem~\ref{runningtime}}]
     We bound the running time of 
     $\alg{v}{K}$ by bounding the number of vertices in the support 
     of $x_{t}$ at each step.  Since $x_{t}$ is at least $\epsilon \norm{}{x_{t}}$
     wherever it is nonzero, we have
     \[ \norm{}{x_{t}}^{2} \geq |\supp(x_{t})| \epsilon^{2} \norm{}{x_{t}}^{2},\]
     and so \[|\supp(x_{t})| \leq \frac{1}{\epsilon^{2}}.\] 

     We can compute $x_{t+1}$ from $x_{t}$ and compute the density of each 
     subgraph $(\set{t}{i},\settwo{t}{j})$ in time proportional 
     to the sum of the degrees of the vertices in $\supp(x_{t})$, 
     which is at most
     \[ O(\Delta |\supp(x_{t})|) = O(\Delta / \epsilon_{t}^{2}) = 
     O(\Delta K^{2} 2^{-2t}).\]
     The total running time is therefore
     \[ 
     \sum_{t=0}^{T} O(\Delta K^{2} 2^{-2t})
     =
     O(\Delta K^{2}).
     \]
 \end{proof}

\section{An approximation algorithm for $d(A)$}\label{S:global} As a
simple application of the techniques developed in the previous
sections, we give an $O(\log n)$-approximation algorithm for the
globally optimum density $d(A)$ by simulating the pruned growth
process for $O(\log n)$ steps.  The algorithm produces a subgraph $(S,T)$
with density $\Omega(d(A) / \log n)$ in time $O(m \log \Delta/d)$,
where $\Delta$ is the maximum degree in the graph, and $d$ is the
average degree.  The algorithm requires $O(\log n)$ passes through the
collection of adjacency lists describing the graph, and requires only $O(n \log\log n)$ bits
of additional storage.  This provides an efficient way to implement
the spectral approximation algorithm of Kannan and Vinay
\cite{Kannan}, which has the same $O(\log n)$ 
approximation guarantee and requires
computing the largest eigenvalue of $A$.

\noindent
\framebox{
\begin{minipage}{.95\textwidth}
    {\noindent $\globalg$:}
    
    Run the following procedure twice with $x_{0} = 1_{L}$ and $x_{0}=1_{R}$:
        \begin{enumerate}
    \item 
        Let $T = \log 2\sqrt{n}$ 
     and $\epsilon_{t} = \frac{2^{t}}{8\sqrt{n}}$.
 \item Compute the pruned growth process vectors $x_{0}, \dots, x_{T}$.
    \item 
    Compute \mbox{$\density(\set{t}{i},\settwo{t}{j})$} for each pair $i$, $j$ and each time $t<T$. 
    \item
    Output the densest subgraph among the sets
    $(\set{t}{i},\settwo{t}{j})$.
\end{enumerate}
\end{minipage}
}

\begin{theorem}\label{powermethod}
    For at least one of the two starting vectors $1_{L}$ and $1_{R}$, there exists a time
     $t \leq T$ and two indices $i$ and $j$ such that the subgraph 
     $(X,Y)$ output by the algorithm satisfies 
\[\density(X,Y) 
\geq \frac{\lambda}{(8 + 4 \log n)} 
\geq \frac{d(A)}{(8+ 4 \log n)}.\] 
\end{theorem}

\begin{theorem}\label{globalrunningtime}
    $\globalg$ runs in time $O(m(1+\log \frac{\Delta}{d}))$,
    where $\Delta$ is the maximum degree in the graph, and $d$ is the average degree.  The algorithm requires $O(n \log \log n)$ bits of additional storage.
\end{theorem}

\begin{proof}[{\bf Proof of Theorem~\ref{powermethod}}]
     Let $\lambda$ be the largest eigenvalue of $A$, and let $\phi$ be an eigenvector with eigenvalue $\lambda$ whose entries are nonnegative. 
     Because $\phi$ is nonnegative, 
     \mbox{$\ip{1_{V}}{\phi} \geq 1$}.
    We will assume that $1_{L}$ has a larger inner product with $\phi$ than $1_{R}$, so that $\ip{1_{L}}{\phi} \geq (1/2)\ip{1_{V}}{\phi} \geq 1/2$.  We let $x_{0} = 1_{L}$, and consider the vectors $x_{0}, \dots, x_{T}$ computed by the algorithm.

We assume 
that 
$\density(\set{t}{i},\settwo{t}{j}) < 
\lambda/8\log(2\Delta/\epsilon_{0}) \leq \lambda/(8+4\log n)$
for every $i$, $j$, and $t \leq T$,
and derive a contradiction.
Under this assumption, Lemma~\ref{growthbound} shows that 
for every $t \leq T$,
     \begin{align*}
         \| x_{t+1} \| 
         &\leq \| \round(x_{t}A) \| \\
         &< 
         \left(2 \log\frac{2\Delta}{\epsilon_{t}} \right)
         \left(\frac{\lambda}{8 \log(2\Delta/\epsilon_{0})} \right) \|x_{t} \|\\
         &\leq \left(\frac{\lambda}{4}\right)\|x_{t} \|.
     \end{align*}
Since $\|x_{0}\| \leq \sqrt{n}$, this implies
\begin{equation}\label{globalgrowthboundeq}
    \| x_{t} \| < \sqrt{n} \left(\frac{\lambda}{4}\right)^{t}\quad
    \mbox{for every $t \leq T$.}
\end{equation}
We will soon prove the following lower bound.
\begin{equation}\label{globallowerboundeq}
    \ip{x_{t}}{\phi} \geq  \ip{1_{L}}{\phi} (\lambda/2)^{t}
\quad \mbox{for every $t \leq T$.}
\end{equation}
When $t=T=\log(2\sqrt{n})$, this will imply
\[\norm{ }{x_{T}} \geq \ip{x_{T}}{\phi} 
\geq \frac{1}{2}(\lambda/2)^{T} 
\geq \sqrt{n}(\lambda/4)^{T},\]
which will contradict equation~\eqref{globalgrowthboundeq},
completing the proof.

We will prove by induction
that equation~\eqref{globallowerboundeq} holds
for every $t \leq T$.  It holds trivially  for $t=0$.  
We define $r_{t}$ to be the vector lost in the
pruning step,
\begin{align*}
    r_{t} 
    &= \round(x_{t-1}A) - x_{t}\\
    &= \round(x_{t-1}A) - \truncate{t}(\round(x_{t-1}A)). 
\end{align*}
The value of $r_{t}$ at any given vertex is at most
$\epsilon_{t} \norm{}{\round(x_{t-1}A)} \leq 2 \lambda \epsilon_{t} \norm{}{x_{t-1}}$.
   Because $\phi$ is nonnegative, 
   $\ip{r_{t}}{\phi} \leq 2 \lambda \epsilon_{t} \norm{}{x_{t-1}} \ip{1_{V}}{\phi}$.
   In fact, we have the slightly stronger statement $\ip{r_{t}}{\phi} \leq 2 \lambda \epsilon_{t} \norm{}{x_{t-1}} \ip{1_{L}}{\phi}$,
 because the support of $r_{t}$ is contained in either $L$ or $R$, and 
 $1_{L}$ has a larger inner product with $\phi$.  We can now bound $\ip{x_{t}}{\phi}$ in terms of $\ip{x_{t-1}}{\phi}$.
\begin{align*}
    \ip{x_{t}}{\phi} 
    &= \ip{\round(x_{t-1}A) - r_{t}}{\phi}\\
    &= \ip{\round(x_{t-1}A)}{\phi} - \ip{r_{t}}{\phi}  \\
    &\geq  \lambda \ip{x_{t-1}}{\phi} - 
    2 \lambda \epsilon_{t} \norm{}{x_{t-1}} \ip{1_{L}}{\phi}.
  \end{align*}
We will assume that the induction hypothesis holds for $t-1$, 
which means that
$\ip{x_{t-1}}{\phi} \geq \ip{1_{L}}{\phi}(\lambda/2)^{t-1}$,
and we have already assumed for the sake of contradiction that
$\| x_{t} \| < \sqrt{n}\left(\lambda/4\right)^{t}$. 
We now show that the induction hypothesis holds for $t$.
\begin{align*}
    \ip{x_{t}}{\phi} 
    &\geq  \lambda \ip{1_{L}}{\phi}(\lambda/2)^{t-1} - 
    2 \lambda \epsilon_{t} 
    \ip{1_{L}}{\phi}\sqrt{n} (\lambda/4)^{t-1}\\
    &\geq  \ip{x_{0}}{\phi}(\lambda/2)^{t} (2 - 8\epsilon_{t}2^{-t}  \sqrt{n})\\
    &\geq  \ip{x_{0}}{\phi}(\lambda/2)^{t}. 
\end{align*}
The last step follows because we have set $\epsilon_{t}$ so that 
\[ \epsilon_{t} = \frac{2^{t}}{8\sqrt{n}}.\]
This completes the proof.
\end{proof}

\begin{proof}[{\bf Proof of Theorem~\ref{globalrunningtime}}]
We can bound the running time of the algorithm by bounding the number of vertices in the support 
of $x_{t}$.  Since $x_{t}$ is at least $\epsilon_{t} \norm{}{x_{t}}$
wherever it is nonzero, we have
\[ \norm{}{x_{t}}^{2} \geq |\supp(x_{t})| \epsilon_{t}^{2} \norm{}{x_{t}}^{2},\]
and so 
\[|\supp(x_{t})| \leq \frac{1}{\epsilon_{t}^{2}} \leq n 2^{-2(t-3)}.\]

We can compute $x_{t+1}$ from $x_{t}$ and compute the density of each 
subgraph $(\set{t}{i},\settwo{t}{j})$ in time proportional to the number of edges incident with 
$\supp(x_{t})$, 
\begin{align*}
    |e(\supp(x_{t}),V)| 
    &\leq 
    \min(m,\Delta |\supp(x_{t})|)\\
    &\leq 
    \min(m,\Delta n2^{-2(t-3)}).
\end{align*}

The total running time over all $T$ steps is at most
\begin{align*}
    \sum_{t=0}^{T} \min(m,\Delta n2^{-2(t-3)})
    &\leq 
    \left(\frac{1}{2}\log (n\Delta/m)+3\right) m 
    +
    \sum_{t=\frac{1}{2}\log (n\Delta/m)+3}^{T} \Delta n2^{-2(t-3)}\\
    &\leq 
    \left( \frac{1}{2}\log (n\Delta/m)+3 \right) m + 2m\\
    &=
    O(m\log (\Delta/d) + m).
\end{align*}

  To bound the amount of space used by the algorithm, notice that storing the vector $x_{t}$ requires \mbox{$n \log \log \frac{1}{\epsilon} = O(n \log \log n)$} bits, 
since each vertex takes one of $\log \frac{1}{\epsilon}$ possible values.
We need only store two vectors at a given time, 
 $x_{t}$ and $\round(x_{t}A)$,
 so the total amount of storage required is $O(n \log \log n)$ bits. 
\end{proof}

 \section{Conclusion} We have shown that it is possible to find a
 dense subgraph near a given vertex without examining the entire
 graph.  The running time of our local algorithm is quadratic in terms
 of the target size $K$, where $K$ must be at least as large as 
 $|S|+|T|$ to produce an approximation of the subgraph $(S,T)$.
 We
 conjecture that a better local algorithm exists.  In particular, it
 would be nice to have an algorithm whose running time depends on the
 size of the subgraph $(X,Y)$ that is produced,
 rather the subgraph $(S,T)$ whose density is approximated.


\begin{thebibliography}{77}	

    \bibitem{Andersen}
R.~Andersen, F.~Chung, and K.~Lang.
\newblock Local graph partitioning using PageRank vectors.
\newblock In {\em Proc. 47th Annual Symposium on Foundations of Computer Science}, (2006).  

    \bibitem{Charikar}
    M.~Charikar.
    Greedy approximation algorithms for finding dense components in a graph. 
    In {\it Proc. Third International Workshop on Approximation Algorithms for Combinatorial Optimization}, (2000).

\bibitem{feige}
U.~Feige, D.~Peleg, and G.~Kortsarz.
The dense k-subgraph problem. 
{\it Algorithmica}, 29(3), 410-421, (2001).

\bibitem{feigeseltser}
U.~Feige and M.~Seltser. On the densest k-subgraph problem.
Weizmann Institute Technical Report CS 97-16, (1997).

\bibitem{GGT}
    G.~Gallo, M.D.~Grigoriadis, and R.~Tarjan.  A Fast Parametric Maximum Flow Algorithm and Applications. 
    In {\it Proc. 39th Annual IEEE Symposium on Foundations of Computer Science}, 370-378 (1998). 

\bibitem{GKT}
D.~Gibson, R.~Kumar, and A.~Tomkins. 
Discovering Large Dense Subgraphs in Massive Graphs.
In {\it Proc. 31st VLDB Conference}, (2005). 

\bibitem{Goldberg}
A.~Goldberg.  
Finding a maximum density subgraph.
Technical report UCB CSD 84/71, University of California, Berkeley,
(1984).

\bibitem{Kannan} 
    R.~Kannan and V.~Vinay.  Analyzing the Structure of Large Graphs.  {\it Manuscript}, (1999).


\bibitem{trawling}
R.~Kumar, P.~Raghavan, S.~Rajagopalan, and A.~Tomkins. 
Trawling the Web for emerging cyber-communities. 
In {\it Proc. 8th WWW Conference}, Computer Networks, 31(11-16):1481-1493, (1999). 


\bibitem{Spielman}
D.~Spielman and S.H.~Teng. 
Nearly-Linear Time Algorithms for Graph Partitioning, Graph Sparsification, and Solving Linear Systems.
In {\it Proc. 36th Annual ACM Symposium on Theory of Computing}, (2004).

\end{thebibliography}
\end{document}